\documentclass[prd,aps,twocolumn,a4paper,showkeys,nofootinbib,showpacs]{revtex4-1}

\usepackage[colorlinks=true,linkcolor=blue,linktocpage=true,anchorcolor=black,citecolor=blue,filecolor=black,menucolor=black,urlcolor=blue]{hyperref}

\usepackage{graphicx,psfrag}
\usepackage{mathrsfs}
\usepackage{amsmath,amsfonts,amssymb}
\usepackage{comment}
\usepackage{ulem}
\usepackage[utf8]{inputenc}
\usepackage[top=2cm,bottom=2cm,left=1.4cm,right=1.4cm]{geometry}

\newcommand{\be}{\begin{equation}}
	\newcommand{\ee}{\end{equation}}
\newcommand{\bea}{\begin{eqnarray}}
	\newcommand{\eea}{\end{eqnarray}}
\newcommand{\bel}{\begin{align}}
	\newcommand{\eel}{\end{align}}

\def\GMc2{G M_{\odot} c^{-2}}

\def\de{\partial}

\def\g{{\gamma}}

\DeclareSymbolFontAlphabet{\mathrsfs}{rsfs}
\DeclareMathAlphabet{\mathcal}{OMS}{cmsy}{m}{n}
\DeclareSymbolFontAlphabet{\mathrsfs}{rsfs}
\DeclareMathAlphabet\mathbfcal{OMS}{cmsy}{b}{n}

\usepackage{color}
\definecolor{cyan}{rgb}{0,0.9,0.9}
\definecolor{orange}{rgb}{0.9,0.5,0}
\definecolor{magenta}{rgb}{1,0,1}
\definecolor{purple}{rgb}{0.8,0.4,0.8}
\definecolor{gray}{rgb}{0.8242,0.8242,0.8242}
\definecolor{dodgerblue}{rgb}{0.12, 0.56, 1.0}

\begin{document}
	
\title{Double copy, Kerr-Schild gauges and the Effective-One-Body formalism}
\author{Anna \surname{Ceresole}${}^{1}$}
\author{Thibault \surname{Damour}${}^{2}$}
\author{Alessandro \surname{Nagar}${}^{1,2}$}	
\author{Piero \surname{Rettegno}${}^{1}$}

\affiliation{${}^1$INFN Sezione di Torino, Via P. Giuria 1, 10125 Torino, Italy} 	
\affiliation{${}^2$Institut des Hautes Etudes Scientifiques, 91440 Bures-sur-Yvette, France}

\begin{abstract}
We look for a classical double copy structure between gravity and electrodynamics by connecting
the descriptions of the scattering of two point masses, and of two point charges, in terms of 
perturbative (post-Minkowskian or post-Lorentzian)  expansions. We do so by recasting available analytical information within the 
effective-one-body formalism using Kerr-Schild  gauges in both cases.
Working at the third perturbative level, we find that the usual linear relation (holding in the probe limit) between the 
adimensionalized electric potential, 
$\tilde{\phi}= \frac{G M}{e_1 e_2} \phi^{\rm el}$, and the Schwarzschildlike gravitational one, $\Phi^{\rm grav}$, 
 is deformed, in the comparable-mass, comparable-charge, case, 
into a {\it nonlinear relation} which becomes {\it universal in the high energy limit}: 
$\Phi^{\rm grav}= 2\tilde{\phi}-5\tilde{\phi}^2 +  18\tilde{\phi}^3$.
\end{abstract}

\maketitle
	
\section{Introduction}
The double copy  is a relation between gauge theory and gravity that is often summarized in 
the motto that ``gravity is the square of gauge theory". 
Such a quadratic relation  appeared in the early development 
of the operator formalism in string theory~\cite{Fubini:1970xj}  through the 
factorized structure of the  graviton vertex operator, $V^{\mu\nu} \propto \partial_z X^\mu  \partial_{\bar z} X^\nu e^{ i k.X}$. At the tree level, a precise expression of gravity 
amplitudes as sums of products of gauge-theory amplitudes was obtained by  Kawai, Lewellen and Tye~\cite{Kawai:1985xq}. Independently of the possible existence of a string theory underlying these
 amplitudes,  Bern et al. ~\cite{Bern:2008qj,Bern:2010yg,Bern:2013yya,Bern:2022wqg} have shown that a quadratic relation
between gauge and gravity amplitudes still holds at the loop level via a duality between
color and kinematics.
 
The {\it quantum origin} of a squaring relation between gravity and gauge theory raises the question
of the possible meaning of such a relation at the {\it classical} level.
Some attempts have been made to find gauges that would exhibit such a squaring relation
at the level of the  classical General Relativity (GR) Lagrangian \cite{Bern:1999ji,Ananth:2007zy,Hohm:2011dz,Cheung:2016say,Ferrero:2020vww,Beneke:2021ilf,Cheung:2022mix}, though this search has not led to
a definitive classical formulation of the double copy idea. Another line of work has been to
exhibit relations between special families of exact solutions in gravity and in gauge theories.
In particular, the role of stationary Kerr-Schild metrics was emphasized in Ref. \cite{Monteiro:2014cda}.
It had been known for a long time \cite{Gurses:1975vu} that the Einstein equations for
metrics satisfying the   Kerr-Schild ansatz \cite{Kerr:2009wic}, namely,
\be \label{KS0}
g_{\mu\nu}(x)= \eta_{\mu\nu} + \Phi(x) k_\mu(x) k_\nu(x)\,,
\ee
where $k_\mu(x)$ is a null vector field (with respect to both the Minkowski background
$\eta_{\mu\nu}$ and $g_{\mu\nu}$) are linear in 
$g^{\mu\nu}(x)= \eta^{\mu\nu} - \Phi(x) k^\mu(x) k^\nu(x)$.  Ref. \cite{Monteiro:2014cda} noted
that, in the case of stationary Kerr-Schild metrics (with $k^0=1$) the corresponding ``single-copy"
gauge potential $A^a_\mu \equiv c^a \Phi(x) k_\mu(x)$ (where $c^a$ is some fixed
Lie-algebra valued element) automatically satisfies Yang-Mills' equations. 
For other works on the classical double copy see Refs. 
~\cite{Luna:2015paa,Kim:2019jwm,Goldberger:2016iau,Carrillo-Gonzalez:2017iyj,Goldberger:2017vcg,Goldberger:2017ogt,Li:2018qap} (see also  Refs. \cite{Adamo:2022dcm,Alawadhi:2023beb}
for reviews of the double copy).

In a different line of research, Refs.~\cite{Damour:2016gwp,Damour:2017zjx,Damour:2019lcq} 
introduced a new approach to the two-body problem in General Relativity based on a direct 
translation of classical or quantum gravitational two-body scattering results into an 
effective-one-body (EOB) \cite{Buonanno:1998gg,Buonanno:2000ef,Damour:2000we}
mass-shell condition, which can be put in the form
\begin{equation} \label{massshell1}
	g_{\rm eff}^{\rm \mu\nu}(\g) P_\mu  P_\nu + \mu^2 =0 \, ,
\end{equation}
where $\mu= \frac{m_1 m_2}{m_1+m_2}$ is the usual effective mass
and  $g^{\rm eff}_{\rm \mu\nu}(\g)$ is an {\it energy-dependent} effective metric encapsulating
the gauge-invariant scattering relation $\chi=\chi(E,J)$ between the scattering angle $\chi$, the two-body
(incoming) angular momentum $J$ and the two-body (incoming) energy $E$. Here, the quantity $\g$ is
linked to the effective energy $E_{\rm eff} =-P_0$  and the total energy $E$ as
\be
 \label{defg}
\g= \frac{E^2-m_1^2-m_2^2}{2m_1 m_2 }=\dfrac{E_{\rm eff}}{\mu} \ .
\ee
The energy dependence of the (inverse) effective metric 
$g_{\rm eff}^{\rm \mu\nu}(\g)$ in Eq.~\eqref{massshell1} replaces
the usual EOB higher-than-quadratic-in-momenta contribution $Q(P,X)$. See also
Ref.~\cite{Cheung:2018wkq} for an alternative way of mapping two-body scattering results
into an effective Hamiltonian.

The scattering approach~\cite{Damour:2017zjx,Cheung:2018wkq} to the classical two-body GR problem 
 prompted the application of modern quantum field theory techniques (and notably of the double
 copy results~\cite{Bern:2008qj,Bern:2010yg}) to the  computation of high-order perturbative
 contributions to the scattering angle, namely at the  3PM
 order~\cite{Bern:2019nnu,Kalin:2020fhe,Damour:2020tta,DiVecchia:2021bdo,Bjerrum-Bohr:2021din} and at the 
 4PM order~\cite{Bern:2021dqo,Bern:2021dqo,Dlapa:2021vgp, Dlapa:2022lmu,Damgaard:2023ttc}.
 This new analytical knowledge was recently successfully compared to numerical relativity
 simulations of the scattering of binary black holes~\cite{Damour:2022ybd,Rettegno:2023ghr}.

In parallel, the old  result, in classical electrodynamics (ED),
 of  Westpfahl~\cite{Westpfahl:1985tsl}  on the scattering angle of two electromagnetically
interacting charges at the second post-Lorentzian (2PL) order (second order in $e^2 \sim e_1 e_2$),
was recently extended to the 3PL order (third order in $e^2$)~\cite{Saketh:2021sri,Bern:2021xze}.
The potential-photon contributions to the scattering have also
been computed at the 4PL and 5PL orders~\cite{Bern:2023ccb}.
Moreover, the scattering angle of two extremal (half-BPS) black holes in ${\cal N}=8$ supergravity
has been computed at the 2PM order in Ref.~\cite{Caron-Huot:2018ape} and (for 
a special case) at the 3PM order
in Refs.~\cite{Parra-Martinez:2020dzs,Herrmann:2021tct,DiVecchia:2021bdo}.

The Kerr-Schild ansatz, Eq.~\eqref{KS0}, is only valid for special classes
of GR solutions, such as isolated black holes, and cannot be applied to the physically
important case of gravitationally interacting binary black holes. This limits the physical
applicability of the simple linear relation 
between $ g_{\mu\nu}(x)= \eta_{\mu\nu} + \Phi(x) k_\mu(x) k_\nu(x)$ 
and, say in ED,  $A_\mu = c \Phi(x) k_\mu(x)$ to the dynamics of
probe objects in stationary external fields. For instance, the classical double-copy 
result of \cite{Monteiro:2014cda} naturally associates, via such a linear relation, 
the external Schwarzschild potential $\Phi= \frac{2 GM_0}{R}$ felt by a test mass $\mu_0$
to the Coulomb potential $A^0 = \frac{e_0}{R}$ felt by a test-charge $e_{\rm test}$.

The aim of this paper is to explore whether a generalization of this double-copy result
exists when considering, within the EOB formalism, the relation between the 
effective metric $g_{\rm eff}^{\rm \mu\nu}(\g)$, Eq. \eqref{massshell1},
describing the scattering of two black holes, and its ED analog, describing
the scattering of two charges. To do so we will, for the first time, formulate
the EOB mass-shell conditions in Kerr-Schild gauges, both in GR and in ED.
Contrary to the usual Kerr-Schild {\it ansatz}, Eq.~\eqref{KS0},
which is a strong restriction on the considered GR metrics, we shall show here that
it is {\it always possible} to cast the effective EOB metric $g_{\rm eff}^{\rm \mu\nu}(\g)$
describing the scattering of two black holes (and its ED analog, introduced below)
in a Kerr-Schild-type {\it gauge}. The use of these gauges will allow us to associate
a gravity potential with an electromagnetic one  in a physical setting going beyond
the interaction of probes with stationary fields, and involving the non stationary
interactions of comparable masses and of comparable charges.

We restrict ourselves here to the 3PM order (including radiation-reaction 
effects)~\cite{Bern:2019nnu,Kalin:2020fhe,Damour:2020tta,DiVecchia:2021bdo,Bjerrum-Bohr:2021din}
because the current treatment of radiative effects at the 4PM order~\cite{Bern:2021yeh,Bern:2021dqo,Dlapa:2021vgp,Dlapa:2021npj,Bini:2022enm,Manohar:2022dea,Dlapa:2022lmu,Bini:2022enm,Dlapa:2023hsl} 
leads to a yet-to-be understood singular behavior of the gravitational scattering 
angle in the high-energy limit. We correspondingly work at the 3PL order in 
ED (including radiation-reaction effects)~\cite{Saketh:2021sri,Bern:2021xze}.

\section{Weak-field expansions and scattering angles}
\label{sec:eob_intro}
Let us first review known 3PM-accurate results (including radiative effects)
for the scattering angle in both GR and ED.
In GR, the PM approximation consists of a formal series in the gravitational constant $G$. 
The scattering angle $\chi$ between two point masses $(m_1,m_2)$ is then 
expressed as a series in the inverse of the angular momentum $J$
 \begin{align}	
\label{eq:chiGR}
\dfrac{\chi^{\rm GR}(E,J)}{2} &=\frac{G m_1 m_2}{J}  \chi_1^{\rm GR}(\g)  +  \left(\frac{G m_1 m_2}{J}\right)^2  \chi_2^{\rm GR}(\g) \nonumber \\
&+  \left(\frac{G m_1 m_2}{J}\right)^3 \chi_3^{\rm GR}(\g) + {O}\left(\frac{G^4}{J^4}\right)\,,
\end{align}
where the $\chi_i^{\rm GR}$ coefficients depend on the incoming energy of the system 
(expressed in terms of $\gamma$, Eq.~\eqref{defg}) and on the masses $m_1, m_2$.
In ED, the expansion parameter is  the numerator $k \, e_1 e_2$ of the Coulomb interaction potential
(which is the electromagnetic analog of $G m_1 m_2$).
The PL-expanded angle reads
\begin{align}	
	\label{eq:chiED}
	\frac{\chi^{\rm ED}(E,J)}{2} &=  \frac{k \,e_1 e_2}{J} \chi_1^{\rm ED}(\g) +  \left(\frac{k\, e_1 e_2}{J}\right)^2 \chi_2^{\rm ED}(\g)  \nonumber \\
	&+  \left(\frac{k\,  e_1 e_2}{J}\right)^3 \chi_3^{\rm ED}(\g)+ { O}\left(\frac{k^4}{J^4}\right)\,,
\end{align}	
where $e_{1,2}$ are the charges of the two bodies. We use units where $c=1$, so that
$\frac{G m_1 m_2}{J}$ and $ \frac{k \,e_1 e_2}{J} $ are dimensionless.
In the following, we generally use Gauss units where $k=1$.

The (dimensionless) coefficients in Eqs.~\eqref{eq:chiGR} and~\eqref{eq:chiED} 
for GR~\cite{Bern:2021dqo,Dlapa:2021vgp,Manohar:2022dea,Dlapa:2022lmu} 
and ED~\cite{Saketh:2021sri,Bern:2021xze,Bern:2023ccb} read
\begin{align}
	\label{eq:chi_i_gr}
	\chi_1^{\rm GR} =&~ \frac{2 \gamma^2 - 1}{\sqrt{\gamma^2-1}}\,, \nonumber \\
	\chi_2^{\rm GR} =&~ \frac{3 \pi}{8} \frac{(5 \gamma^2 - 1)}{h}\,, \nonumber \\
	\chi_3^{\rm GR} =&~ \frac{64 \gamma^6 -120 \gamma^4 + 60 \gamma^2 - 5}{3 p_\infty^3} \nonumber \\
	&- \frac{\nu}{h^2 p_\infty^3}\Bigg\{\frac{4}{3}\gamma\, p_\infty^4\left(25+14\gamma^2\right)\nonumber \\
	&+\frac{ 2 p_\infty}{3}\left(2\gamma^2-1\right)^2\left(5\gamma^2-8\right) \nonumber \\
	&-4\Big[\gamma\left(2\gamma^2-3\right)\left(2\gamma^2-1\right)^2\nonumber \\
	&+2p_\infty^3\left(3+12\gamma^2-4\gamma^4\right)\Big]{\rm arcsinh}\sqrt{\dfrac{\gamma-1}{2}}\Bigg\}\,,
\end{align}
and 
\begin{align}
		\label{eq:chi_i_em}
		\chi_1^{\rm ED} =&~ -\frac{\gamma}{\sqrt{\gamma^2-1}}\,, \nonumber \\
		\chi_2^{\rm ED} =&~  \frac{\pi}{4 h}\,, \nonumber \\
		\chi_3^{\rm ED} =&~ \frac{1}{3 h^2 p_\infty^3}\Bigg\{3\gamma-2\gamma^3\left(1+3\nu p_\infty\right)\nonumber \\
		&+\left(h^2-1\right)\left(\gamma^3+3\gamma^2-3\right) \nonumber \\
		&+12\nu \gamma^2 {\rm arcsinh}\sqrt{\dfrac{\gamma-1}{2}}\nonumber \\
		& + 2 \nu \gamma^2 p_\infty^3 \left(\frac{e_1}{m_1}\frac{m_2}{e_2} +\frac{e_2}{m_2}\frac{m_1}{e_1}\right)\Bigg\} \ ,
\end{align}
where $\nu \equiv \mu/M$ is the symmetric mass ratio of the system, $M=m_1+m_2$, $p_\infty \equiv \sqrt{\gamma^2 - 1}$
and $h$ the mass-rescaled total energy of the system, $h \equiv E/M=\sqrt{1+2\nu\left(\gamma-1\right)}$.

\section{Effective one body formalism and Kerr-Schild gauges}
\label{eob:KS}
The original formulation of the EOB formalism~\cite{Buonanno:1998gg,Buonanno:2000ef,Damour:2000we} 
was aimed at describing the dynamics of gravitational bound states. The Hamiltonian of the
binary system was then defined by solving (for the effective energy $E_{\rm eff}=- P_0$,
linked to the real energy via Eq.~\eqref{defg})
a mass-shell condition of the general form 
$g_{\rm eff}^{ \mu\nu} P_\mu  P_\nu + \mu^2 +Q(X,P)=0$, with 
 $Q(X^\mu,P_\mu)$  accounting for higher than quadratic-in-momenta contributions. When generalizing the EOB formalism to the case
of scattering states \cite{Damour:2016gwp,Damour:2017zjx,Damour:2019lcq,Bini:2020nsb} it was found
convenient to use (as is always possible) an {\it energy gauge} where  $Q(X^\mu,P_\mu)$ is
a function only of $\g$ and $R= |X^i|$ (and where $g_{\rm eff}^{ \mu\nu}$ is a Schwarzschild metric
of mass $M=m_1+m_2$). When using such an EOB energy gauge the PM
expansion of $Q(X,P)=G^2 Q_2(\g)/R^2+ G^3 Q_3(\g)/R^3+\cdots$ can be absorbed in a redefinition
of the time-time component of $g_{\rm eff}^{ 00}$.  This redefinition leads to a simple-looking mass-shell condition 
of the type of Eq. \eqref{massshell1} involving a $\g$-dependent effective metric 
$g_{\rm eff}^{ \mu\nu}(\g,\nu)$ that is a deformation
(ruled by the parameter $\nu$) of the Schwarzschild metric. 
The test-mass limit $\nu \to 0$ of $g^{\rm eff}_{ \mu\nu}(\g,\nu)$ thus reduces to 
the  Schwarzschild metric $ds^2=-(1-2GM/R) dt^2+ dR^2/(1-2GM/R)+ R^2(d\theta^2+\sin^2 \theta d\phi^2)$.  

It is well known that the  Schwarzschild metric can be put, by changing the time coordinate 
$t$ into $T= t- 2GM \log(\frac{R}{2GM}-1)$
into the Kerr-Schild form 
$ds^2=- dT^2 + dR^2+ R^2(d\theta^2+\sin^2 \theta d\phi^2)+ \frac{2GM}{R} (dT-dR)^2$, whose
Cartesian-coordinates version reads (with  $k_\mu= (-1, \frac{X^i}{R})$)
\be 
\label{KSSchw}
g^{\rm Schw}_{\mu \nu}=\eta_{\mu \nu}+  \frac{2GM}{R} k_\mu k_\nu \ .
\ee

When considering the comparable-mass case $\nu \neq 0$, it is not enough to use a coordinate transformation
to put the  post-Schwarzschild EOB effective metric  $g^{\rm eff}_{ \mu\nu}(\g,\nu)=g_{\mu\nu}^{\rm Schw}+O(\nu)$ 
into a Kerr-Schild form. However, one of the defining features of the EOB approach is to make use of the much larger class
of (time-independent) canonical transformations, which notably leave invariant the gauge-invariant 
scattering function $\chi=\chi(E,J)$.  It can be shown that, by using a suitable canonical transformation,
the phase-space mass-shell condition Eq. \eqref{massshell1} can be put into a ``canonical Kerr-Schild gauge"
which has the same formal expression, namely
\begin{equation} \label{massshell2}
	g_{\rm eff}^{\rm \mu\nu}(\g, \nu) P_\mu  P_\nu + \mu^2 =0 \, ,
\end{equation}
with an effective metric of the Kerr-Schild form, i.e., in Cartesian coordinates,
$g^{\rm eff}_{\mu\nu} = \eta_{\mu\nu} + \Phi(R, \g, \nu) k_\mu k_\nu$, with inverse
\bea \label{KS1}
g_{\rm eff}^{\mu\nu} &=& \eta^{\mu\nu} - \Phi(R, \g, \nu) k^\mu k^\nu.
\eea
Here $k_\mu= (-1, \frac{X^i}{R})$ and $k^\mu= (1, \frac{X^i}{R})$ describe an outgoing\footnote{By
time reversal one could also work with an ingoing $k^\mu$.} null congruence centered on the origin.

The simplest way to determine the PM expansion of the so-defined Kerr-Schild gravity 
potential\footnote{Starting at the third perturbative order this potential includes
radiative effects from the corresponding radiation-reacted scattering angle~\cite{Damour:2022ybd}.} 
$ \Phi(R, \g, \nu)$ is to identify the PM expansion of the scattering angle defined by the
mass-shell condition Eq.~\eqref{massshell2} to the known PM expansion of the function  $\chi=\chi(E,J)$.
[The considered canonical transformation leaves invariant both $E_{\rm eff}=- P_0$ and $J= P_\phi$.]
Considering motions in the equatorial plane $\theta=\frac{\pi}{2}$), the scattering angle is given by 
\begin{equation}
\label{eq:chi_int}
	\chi + \pi = -\int_{-\infty}^{+\infty} \dfrac{\de P_R(R,J,\gamma,\nu)}{\de J} dR \ ,
\end{equation}
where the function $ P_R(R,J,\gamma,\nu)$ is obtained by solving the mass-shell condition 
Eq.~\eqref{massshell2}, and where the limits of integration on the radial variable $R=\mp \infty$ 
denote the incoming and final states (at $t=\mp \infty$).
	
\section{Application to General Relativity}
\label{sec:GR}
	
Let us focus first on the  GR case, i.e. the scattering of two (non charged) black holes.
It is convenient to use suitably mass-rescaled variables, $x^i\equiv r n^i =\frac{X^i}{GM}$,
$p_i\equiv \frac{P_i}{\mu}$,  $- p_0 =\frac{E_{\rm eff}}{\mu}= \g$, $j =\frac{J}{G M \mu}=\frac{J}{G m_1 m_2}$
and the dimensionless Newtonianlike potential $ u\equiv \frac{GM}{R}= \frac{1}{r}$. 
The rescaled Kerr-Schild-gauge mass-shell condition rewrites as
\begin{equation}
	\left(\eta^{\mu\nu}-\Phi k^\mu k^\nu\right) p_\mu p_\nu + 1 =0 \, ,
\end{equation}
and yields
$p_{r} = (-\gamma \, \Phi  \pm \sqrt{\gamma^2-(1-\Phi)\left(1+ j^2 u^2\right)})/(1-\Phi)$,
where the sign $\pm$ changes between the incoming and outgoing parts of the hyperboliclike motion.
The scattering angle is then obtained from the equation
\begin{equation}
	\label{eq:chi_gr}
	\pi + \chi(\gamma,j)=2 j \int_0^{{u_{\rm max}}(\gamma,j)}\dfrac{du}{\sqrt{\gamma^2 - (1-\Phi)\left(1+j^2 u^2\right)}} \, ,
\end{equation}
where $u_{\rm max}(\gamma,j)=(G M)/r_{\rm min}(\gamma,j)$, with $r_{\rm min}(\gamma,j)$ 
denoting the turning point of the scattering orbit.	
The coefficients of the PM-expanded potential
$\Phi(u,\g,\nu) = \Phi_1(\gamma) u +  \Phi_2(\gamma) u^2 +\Phi_3(\gamma) u^3 +  O\left(u^4\right),$
are  determined by PM-expanding the integrand of Eq.~\eqref{eq:chi_gr} (taking the {\it partie finie} of 
the resulting divergent integrals~\cite{Damour:1988mr}), and comparing the result to Eq.~\eqref{eq:chi_i_gr}. 
The $\Phi_i(\gamma)$'s up to 3PM read
\begin{widetext}
	\begin{align}
	\label{eq:phi_i_gr}
			\Phi_1 & =  2 \, , \nonumber \\
			\Phi_2 & = -\left(1 - \dfrac{1}{h}\right)\dfrac{3\left(5\gamma^2-1\right)}{3\gamma^2-1} \,  , \nonumber \\		
\Phi_3 & = \left(1-\dfrac{1}{h}\right)\dfrac{9 \left(5\gamma^2-1\right)\left(8\gamma^4-8\gamma^2+1\right)}{\left(4\gamma^2-1\right)\left(3\gamma^2-1\right)p_\infty^2}-\dfrac{2 \nu}{\left(4\gamma^2-1\right) h^2}\left[2\gamma\left(25+14\gamma^2\right)+\dfrac{\left(5\gamma^2-8\right)\left(2\gamma^2-1\right)^2}{p_\infty^3}\right]\nonumber \\
			&+\dfrac{12 \nu}{\left(4\gamma^2-1\right) p_\infty^4 h^2}\left[\left(6+24\gamma^2-8\gamma^4 \right)p_\infty^3+ \gamma\left(2\gamma^2-3\right)\left(2\gamma^2-1\right)^2\right]{\rm arcsinh}\sqrt{\dfrac{\gamma-1}{2}} \, .    
		\end{align}
\end{widetext}
Noting the presence of explicit factors $\nu$, and since $h\to 1$ as $\nu \to 0$, 
we see that the probe limit of $\Phi(u,\g,\nu)$ is the Schwarzschild Kerr-Schild value
$\Phi(u,\g,\nu=0)= 2 u=  \frac{2 GM}{R}$, consistently with Eq.~\eqref{KSSchw}.
An analogous procedure can be applied to ${\cal N}=8$ supergravity,
see Appendix~\ref{sec:sugra}.
		
\section{Application to electrodynamics}
\label{sec:ED}
The reduction of the two-body problem in ED (at the $\frac1{c^4}$ level)
to an EOB description  was first studied 
in Ref.~\cite{Buonanno:2000qq}, within the bound-state framework used at the
time in the corresponding GR case~\cite{Buonanno:1998gg,Buonanno:2000ef}.
One of the possible EOB mappings considered in~\cite{Buonanno:2000qq}
used an energy-dependent scalar potential $A^0(R, \g,\nu)$, in a static gauge
$A^i(R, \g,\nu)=0$.

 Here we focus instead only on scattering motions at the third order in $e^2 \sim e_1 e_2$,
 and at all orders in $\frac{v}{c}$, and, by analogy  with the GR case
 we introduce a single-copy version of the  Kerr-Schild-like gauge for the electromagnetic
 4-potential entering the ED analog of the EOB GR mass-shell condition Eq.~\eqref{massshell2},
 \begin{equation}
	\label{massshellED}
	\eta^{\mu\nu}(P_\mu- {\cal A}^{\rm eff}_\mu)(P_\nu- {\cal A}^{\rm eff}_\nu)+\mu^2=0 \, ,
\end{equation}
i.e. an electromagnetic 4-potential of the form \hbox{${\cal A}^{\rm eff}_\mu\equiv \phi^{\rm el} k_\mu$},
with the same $k_\mu = (-1, \frac{X^i}{R})$ as in the GR 
case\footnote{Similarly to its GR counterpart, $\phi^{\rm el}$ 
incorporates radiative effects starting at the 3PL level.}.
Here $ {\cal A}^{\rm eff}_\mu = e_{\rm eff} { A}^{\rm eff}_\mu$ incorporates the charge $e_{\rm eff}$
of the considered effective charged particle moving in the effective external 4-potential ${ A}^{\rm eff}_\mu$.

We recall that the single-copy version of a Schwarzschild black hole with Kerr-Schild potential  
$\Phi= \frac{2 GM_0}{R}$ is  $A^\mu =\frac{e_0}{R} k_\mu= c \Phi k_\mu= $,  with $c=\frac{e_0}{2GM_0}$.
This result naturally associates the GR dynamics of a test-mass $\mu_0 \ll M_0$ around
a Schwarzschild black hole with the dynamics of a test charge $e_{\rm test} \ll e_0$ in the external 4-potential
 $A_\mu =\frac{e_0}{R} k_\mu$. Our aim here is to see whether this linear probe-limit association still
 exists when considering comparable masses on one side, and comparable charges, on the other side.
 
 As in the GR case,we work with the rescaled variables
$p_i= \frac{P_i}{\mu}$, $x^i = \frac{X^i}{GM}$, 
 $\g = - \frac{P_0}{\mu}$, $j=\frac{J}{G m_1 m_2}$, and $u = \frac{GM}{R}= \frac1{r}$.
 In addition, we rescale the effective electromagnetic interaction potential $\phi^{\rm el}={\cal A}^0_{\rm eff}=e_{\rm eff}A^0_{\rm eff}$
 (with dimensions of electron-volt) into the dimensionless  potential
 $\phi \equiv \phi^{\rm el}/\mu$.  
 Neither $\phi^{\rm el}$ nor $\phi$ involve Newton's constant $G$. 
 Beware, however, that our other rescalings introduce $G$, in some
 of the variables describing a purely ED system. The rescaled ED 
 mass-shell condition reads
\begin{equation}
	\label{eq:mass_shell}
	\eta^{\mu\nu}(p_\mu-\phi k_\mu)(p_\nu- \phi k_\nu)+1=0 \, ,
\end{equation}
yielding
$p_r = \phi \pm \sqrt{\phi^2+2\phi\gamma+ \gamma^2- j^2/r^2-1}$, 
and the EOB scattering angle:
\begin{equation}
	\label{eq:chiED_bis}
	\pi + \chi(\gamma,j) = 2j\int_0^{u_{\rm max}(\gamma,j)}\dfrac{du}{\sqrt{\gamma^2-1-j^2 u^2 +\phi\left(2\gamma +\phi\right)}} \, .
\end{equation}
The (dimensionless) electric potential $\phi \equiv \phi^{\rm el}/\mu$ is PL-expanded as
$\phi = \phi_1(\gamma) u +  \phi_2(\gamma) u^2 +\phi_3(\gamma) u^3 +  O\left(u^4\right)$.
The (dimensionless) $\phi_n$ coefficients are determined by comparing the PL-expansion of Eq.~\eqref{eq:chiED_bis} to Eq.~\eqref{eq:chi_i_em}. We express  the $\phi_n$'s in terms
of the (Papapetrou-Majumdar~\cite{Papaetrou:1947ib,Majumdar:1947eu} inspired)
dimensionless rescaled charges ($i=1,2$) 
$\hat{e}_i \equiv e_i/( \sqrt{G}\,  m_i) ,$
\begin{align}
\label{eq:phi_i_em}
		\phi^{}_1 &= \hat{e}_1 \hat{e}_2\, , \\
		\phi^{}_2 &= \left(\hat{e}_1 \hat{e}_2\right)^2\dfrac{1}{2\gamma}\left(1-\dfrac{1}{h}\right)\, , \\
		\phi^{}_3 & = \left(\hat{e}_1 \hat{e}_2\right)^3 \Bigg\{\dfrac{2\gamma^2-1}{2\gamma^2 p_\infty^2}\left(1-\dfrac{1}{h}\right)+\dfrac{\nu}{\gamma h^2}\left(\dfrac{\gamma^3}{p_\infty^3}-1\right)\nonumber \\
&-\dfrac{\gamma \nu}{3 p_\infty h^2}\left[2+\frac{\left(\hat{e}_1-\hat{e}_2\right)^2}{\hat{e}_1\hat{e}_2}\right]- 
\dfrac{2 \gamma \nu}{p_\infty^4 h^2}{\rm arcsinh}\sqrt{\dfrac{\gamma-1}{2}}\Bigg\} \, ,
\end{align}
where each PL order corresponds to a power of $  \hat{e}_1 \hat{e}_2$.

\begin{figure}[t] 
\centering\includegraphics[width=.45\textwidth]{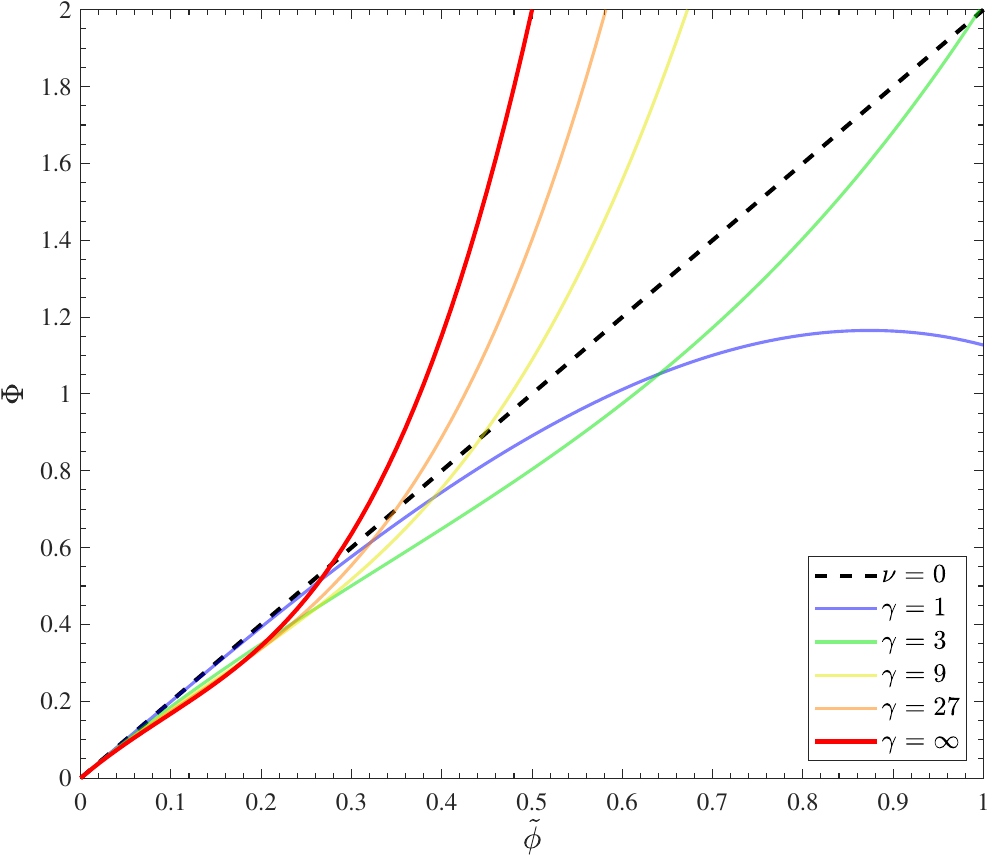}
\caption{GR EOB Kerr-Schild-gauge potential $\Phi$ as a function of the rescaled ED EOB Kerr-Schild-like  potential $\tilde{\phi}$ at the third (post-Minkowskian or post-Lorentzian) perturbative level.
The dashed black line represents the probe limit, where there is a simple, linear relation between the 
Kerr-Schild Schwarzschild potential and the Coulomb one. The red line marks the high-energy limit, 
where the  GR EOB Kerr-Schild potential exhibits a universal non-linear double-copy-type dependence 
on the corresponding ED EOB Kerr-Schild one. The other lines correspond to different energies, 
$\gamma=\{1,3,9,27\}$. For these, we set $\nu=1/4$ and $\hat{e}_1 = \hat{e}_2 = 1$. } 
\label{fig:phi}
\end{figure}

\section{Search for double copy in the  Kerr-Schild-gauge EOB potentials}
\label{sec:DC}
In the probe limit  $\frac{m_1}{m_2} \to 0$, and $\frac{e_1}{e_2} \to 0$, with, say,
 $\frac{e_1}{m_1}$ fixed, the Kerr-Schild-gauge EOB gravitational and
 electromagnetic potentials have the following limits: 
 $\Phi \to 2 u= \frac{2GM}{R}$ and $\phi \to \hat{e}_1 \hat{e}_2 u = \frac{e_1 e_2}{\mu R}$.
 As already said, and emphasized in Ref.~\cite{Monteiro:2014cda}, the probe limit leads to a simple linear relation between the two types
 of Kerr-Schild-gauge potentials, namely
 \be \label{linDC}
 \Phi \overset{\rm probe \, limit}{=} \frac{2 G m_1 m_2}{e_1 e_2} \phi= \frac{2 G M}{e_1 e_2} \phi^{\rm el}.
 \ee
 For the comparable-mass and comparable-charge case, our results above show that 
 the linear double-copy relation Eq.~\eqref{linDC} becomes deformed into a  {\it  nonlinear double-copy relation} 
 between $\phi$ and $\Phi$. Its expression is simplified by introducing a rescaled version of $\phi$, namely
$\tilde{\phi} \equiv \frac{\phi}{\hat{e}_1\hat{e}_2}= \frac{G m_1 m_2}{e_1 e_2} \phi=\frac{G M}{e_1 e_2} \phi^{\rm el}$.
Eliminating $u$ between the PM expansion of $\Phi$ and the PL expansion of $\tilde{\phi}$, we get
the following explicit nonlinear relation that we henceforth truncate at the third perturbative order
\begin{widetext}
\begin{align}
\label{eq:nonlinear_dc}
		\Phi\left(\tilde{\phi}; \g, \nu, \hat e_1, \hat e_2\right) &= 2\tilde{\phi} - \left(1-\frac{1}{h}\right)\left[\frac{\hat{e}_1\hat{e}_2}{\gamma} + \frac{3\left(5\gamma^2-1\right)}{\left(3\gamma^2-1\right)}\right]\tilde{\phi}^2 + \Bigg\{9\left(1-\frac{1}{h}\right)\frac{\left(5\gamma^2-1\right)\left(8\gamma^4-8\gamma^2+1\right)}{\left(3\gamma^2-1\right)\left(4\gamma^2-1\right)p_\infty^2} \nonumber \\
		&- 2\nu\frac{\left(5\gamma^2-8\right)\left(2\gamma^2-1\right)^2+2\gamma p_\infty^3\left(25+14\gamma^2\right)}{h^2 p_\infty^3 \left(4\gamma^2-1\right)} +\left[\frac{3\left(1-h^2\right)\left(5\gamma^2-1\right)}{h^2 \gamma \left(3\gamma^2-1\right)}+\frac{2\gamma\nu}{3 h^2 p_\infty}\left(\hat{e}_1-\hat{e}_2\right)^2\right]\hat{e}_1\hat{e}_2 \nonumber \\
		&+ \left[\frac{-3p_\infty \left(1-h\right) + 2\nu\gamma^3\left(\gamma-2\right)\left(2\gamma+1\right)+6\nu\gamma p_\infty \left(\gamma-1\right)}{3 h^2 \gamma^2 p_\infty^3}\right]\hat{e}_1^2\hat{e}_2^2\nonumber \\
		&+\frac{4\nu}{h^2 p_\infty^4}\left[3\frac{\gamma\left(2\gamma^2-3\right)\left(2\gamma^2-1\right)^2-2p_\infty^3\left(4\gamma^4-12\gamma^2-3\right)}{4\gamma^2-1}+\gamma \hat{e}_1^2 \hat{e}_2^2\right]{\rm arcsinh}\sqrt{\dfrac{\gamma-1}{2}}
		\Bigg\}\tilde{\phi}^3 \ .
\end{align}
\end{widetext}
We propose to view this functional link (truncated at the third order) as a nonlinear double-copy relation between gravity and gauge theory,
in the context of the scattering of comparable masses, and comparable charges. It is a nonlinear
deformation (ruled by $\nu$) of the simple probe-limit double-copy
relation Eq.~\eqref{linDC}, to which Eq.~\eqref{eq:nonlinear_dc} reduces when $\nu\to 0$.
The nonlinear double-copy relation Eq.~\eqref{eq:nonlinear_dc} is illustrated in Fig.~\ref{fig:phi}
for various values of $\g$. For simplicity, we only display the probe limit $\nu = 0$,
and the equal-mass case, $\nu = \frac14$. Concerning the dependence on $\hat e_1$ and $\hat e_2$,
we also consider for simplicity only the case  $\hat e_1=\hat e_2=1$. For most
energies, the dependence of the nonlinear link Eq.~\eqref{eq:nonlinear_dc} on $\hat e_1$ and $\hat e_2$,
is rather mild. There is, however, an exception when considering the low-velocity limit $\g =\frac{1}{\sqrt{1-v^2}}\to 1$.
In this limit, the presence of radiative contributions to the ED scattering angle (linked
to Larmor dipolar radiation) entails a $v^{-1}$ behavior in the gauge potential $\phi$,
which then causes  a $v^{-1}$ behavior in the $\Phi(\phi)$ relation. See Appendix~\ref{sec:sugra}
 for a discussion.

It is remarkable that all the parameter dependence entering the relation 
Eq. \eqref{eq:nonlinear_dc} disappears when considering the high-energy limit, $\gamma \rightarrow +\infty$.
When $\gamma \rightarrow +\infty$ the nonlinear relation Eq.~\eqref{eq:nonlinear_dc}
simplifies into the following {\it universal}, but {\it nonlinear}, link
\begin{equation}
\label{eq:Phi_he}
	\Phi\left(\tilde{\phi}\right) \overset{\gamma \rightarrow +\infty}{=} 2 \tilde{\phi} -5 \tilde{\phi}^2 + 18\tilde{\phi}^3\,.
\end{equation}
The universal curve Eq.~\eqref{eq:Phi_he} is highlighted as a thick red line in Figure~\ref{fig:phi}.
Other curves, for increasing $\g$'s, illustrate the emergence of the universal
high-energy limit Eq.~\eqref{eq:Phi_he}.

In view of its simplicity and universality, the high-energy result of Eq.~\eqref{eq:Phi_he} is
the most important one in our search for a double-copy relation,  beyond the probe limit,
in the two-body problem. It is crucial for obtaining this link to use the EOB Kerr-Schild gauges 
introduced above. The factorization of two $k_\mu$'s in the effective metric, and of one $k_\mu$ in the effective gauge potential,
corrects for the well-known fact that, in the high-energy limit, gravity couples to the square of energy
while gauge theory couples to the first power of energy. This allows the corresponding Kerr-Schild
potentials to have finite limits as $\gamma \rightarrow +\infty$. Actually, the existence
of the limiting result Eq.~\eqref{eq:Phi_he} is the consequence of another remarkable property
of our Kerr-Schild potentials in the high-energy limit. Indeed, when $\g \to \infty$,
$\Phi(R;\g)$ and $\phi(R;\g)$ have the following finite limits (when truncating at third order):
\be \label{HEPhi}
\Phi(R;\g) \overset{\gamma \rightarrow +\infty}{=} 2 u - 5 u^2 + 18 u^3  \,,
\ee
and
\be
\phi(R;\g) \overset{\gamma \rightarrow +\infty}{=} \hat e_1 \hat e_2\, u \,.
\ee
Note also that (see Appendix~\ref{sec:sugra}) the high-energy
limit of the ${\cal N}=8$ supergravity Kerr-Schild potential coincides with
the GR one (as expected in view of the high-energy universality of the gravitational
scattering angle~\cite{DiVecchia:2020ymx,Bern:2020gjj}).

The existence of these finite high-energy limits delicately depends
on the definition of our Kerr-Schild potentials. In particular,  
the 3PL, $O(e^6)$, contribution to the ED scattering angle, Eq.~\eqref{eq:chi_i_em},
does not lead to a finite limit when $\gamma \rightarrow +\infty$
at fixed angular momentum (indeed $\chi_3^{\rm ED}\to +\infty$), 
while the corresponding 3PL potential contribution vanishes: $\phi_3 \to 0$. 
For a previous discussion of the high-energy limit of the EOB 
dynamics see~\cite{Damour:2017zjx}.

\section{Conclusions}
\label{sec:conclusions}
	
We looked for hints of a {\it classical} double-copy structure between gravity and electrodynamics
beyond the existing results, which are essentially limited to the dynamics of probes in 
stationary external fields. 
We compared the classical scattering of two gravitationally interacting comparable masses
to that of two electromagnetically interacting comparable charges.
We introduced a new way of translating, within an effective-one-body approach, the gauge-invariant
information contained in the scattering function, $\chi(E,J)$. 
This new way consists in using, on the gravity
side, a Kerr-Schild gauge [parametrized by a single effective radial gravity potential $\Phi(R)$]
for the effective metric encoding the gravity scattering function, and, on the
electrodynamics side, a single-copy version of the Kerr-Schild gauge [parametrized
by a single electric potential $\phi(R)$] for the effective 4-potential
encoding the gauge scattering function. 
Working at the third perturbative order, we showed that there exists a nonlinear functional link ${\cal F}$ between these two potentials: 
$\Phi={\cal F}(\phi)$, Eq. \eqref{eq:nonlinear_dc}. In the probe limit ${\cal F}$ reduces to the usual linear 
double-copy relation between gravity and gauge theory, Eq. \eqref{linDC}, appearing when considering  single sources
(massive or charged) in Kerr-Schild gauges \cite{Monteiro:2014cda}.
We propose to view  ${\cal F}$ as a {\it nonlinear double-copy} relation between gravity and gauge theory,
in the context of the scattering of comparable masses, and comparable charges. This nonlinear
double-copy relation  generally depends on several parameters (energy, mass ratio, charge-to-mass ratios),
but becomes {\it universal} in the high-energy limit,  Eq. \eqref{eq:Phi_he}. Our results open new avenues for exploring classical versions of the double-copy. They also suggest new ways of applying the effective-one-body
approach to the description of the gravitational interaction of two black holes.

\begin{acknowledgments}
P. ~R. and A.~C. thank the hospitality and the stimulating environment of the Institut des Hautes Etudes Scientifiques. We thank Stefano De Angelis and Rafael Porto for useful comments.
P.~R. is supported by the Italian Minister of University and Research (MUR) via the 
PRIN 2020KB33TP, {\it Multimessenger astronomy in the Einstein Telescope Era (METE)}.
The present research was also partly supported by the ``\textit{2021 Balzan Prize for 
Gravitation: Physical and Astrophysical Aspects}'', awarded to Thibault Damour.
\end{acknowledgments}

\appendix
\section{${\cal N}=8$ supergravity}
\label{sec:sugra}
As a complement, we can apply our GR treatment to
the scattering  of two (half-BPS) extremal black holes in ${\cal N}=8$ supergravity. These 
systems are generally parametrized by three angles~\cite{Caron-Huot:2018ape}. In one particular case
(a single angle, taken to be $\phi=\frac{\pi}{2}$), the scattering angle has been computed 
to $O(G^3)$~\cite{Parra-Martinez:2020dzs,Herrmann:2021tct}, and reads
\begin{align}
	\label{eq:chi_i_sugra}
	\chi_1^{\rm sugra} &= \frac{2 \gamma^2}{\sqrt{\gamma^2-1}}\,, \nonumber \\
	\chi_2^{\rm sugra} &= 0\,, \nonumber \\
	\chi_3^{\rm sugra} &= -\frac{8 \gamma^6}{3 p_\infty^3} 
	+\frac{16 \nu \gamma^4}{h^2 p_\infty^3}\Bigg[p_\infty \gamma^2 \nonumber \\
	&- 2\left(p_\infty^3+2\gamma-\gamma^3\right){\rm arcsinh}\sqrt{\dfrac{\gamma-1}{2}}\Bigg]\,.
\end{align}
The scattering of these systems involves, besides the exchange of the graviton, $h_{\mu \nu}$,
the exchange of two graviphotons, say $A_\mu$, and $B_\mu$, and four scalar fields (including the universal
dilaton), say $\Phi, A, B, C$~\cite{DiVecchia:2021ndb} (though the fourth scalar field, $C$, does not couple 
when $\phi=\frac{\pi}{2}$). This situation  suggests that a natural EOB description of the scattering
might involve a mass-shell condition of a more general type than the usual GR one, say, at least
\begin{equation} \label{massshell3}
G_{\rm eff}^{\rm \mu\nu} (P_\mu - {\cal A}^{\rm eff}_\mu)( P_\nu  - {\cal A}^{\rm eff}_\nu)+ \mu^2 =0 \, ,
\end{equation}
with an effective metric $G_{\rm eff}^{\rm \mu\nu}$  incorporating the scalar exchanges.
For simplicity, we, however, use a standard GR-type mass-shell condition (without
${\cal A}^{\rm eff}_\mu$ coupling). Using a Kerr-Schild gauge for the effective metric, i.e.,
$G_{\rm eff}^{\rm \mu\nu}= \eta^{\mu\nu}- \Phi^{\rm sugra} k^\mu k^\nu$, and PM-expanding
the supergravity potential, $\Phi^{\rm sugra}= \Phi^{\rm sugra}_1 u +  \Phi^{\rm sugra}_2 u^2+
 \Phi^{\rm sugra}_3 u^3 + O(u^4)$, leads to the values
\begin{align}
	\label{eq:phi_i_sugra}
	\Phi^{\rm sugra}_1 & = \dfrac{4\gamma^2}{2\gamma^2-1} \, , \nonumber \\
	\Phi^{\rm sugra}_2 & = -\dfrac{12\gamma^4}{(2\gamma^2-1)^2}\dfrac{5\gamma^2-1}{3\gamma^2-1}\,,\nonumber\\
	\Phi^{\rm sugra}_3 & =\dfrac{24\gamma^6}{4\gamma^2-1}
	\Bigg\{\dfrac{1-12\gamma^2+48\gamma^4}{\left(2\gamma^2-1\right)^3\left(3\gamma^2-1\right)} \nonumber \\
	&+\dfrac{2\nu}{h^2 p_\infty^3} \left[1+2\dfrac{\gamma\left(\gamma^2-2\right)-p_\infty^3}{\gamma^2 p_\infty}{\rm arcsinh}\sqrt{\dfrac{\gamma-1}{2}}\right]\Bigg\}\,.
\end{align}
Consistently with the known universality of high-energy gravitational 
scattering~\cite{DiVecchia:2020ymx,Bern:2020gjj}, the ultrarelativistic limit, $\gamma\to\infty$, 
of $\Phi^{\rm sugra}$ coincides with the GR one Eq.~\eqref{HEPhi}.
By contrast, the  low-velocity limit of $\Phi^{\rm sugra}$ differs from that of GR 
notably because of  the low-velocity behavior of radiative effects entering at the 3PM level.

The low-velocity expansion, $v \to 0$ (with $\gamma \equiv \frac1{\sqrt{1-v^2}}$),
 of the 3PM coefficient of the Kerr-Schild GR potential reads
$\Phi_3 \overset{v \to 0}{=} -5\nu + \frac{8 \, \nu}{5}\, v + O(v^2)$.
By contrast, the low velocity expansion of  $\Phi^{\rm sugra}$ entails 
a $v^{-1}$ behavior in  the $O(G^3)$ coefficient $\Phi^{\rm sugra}_3$, namely 
\hbox{$\Phi^{\rm sugra}_3 \overset{v \to 0}{=} + \frac{80 \,\nu}{3} v^{-1} + O(v^0)$}.
There is an analogous low-velocity behavior in the 3PL coefficient of the Kerr-Schild ED potential
when $\hat e_1 \neq \hat e_2$,
$\phi_3 \overset{v \to 0}{=} -\frac{\nu}{3}\hat{e}_1^2\hat{e}_2^2\left(\hat{e}_1-\hat{e}_2\right)^2 v^{-1} +O(v^0)$.
The presence of odd powers of $v$ in the post-Newtonian
expansions of the potentials $\Phi$, $\Phi^{\rm sugra}$, and $\phi$ arises because these effective
potentials were deduced from the radiation-reacted values of the corresponding scattering angles.
The $O(v^1)$ term in $\Phi_3$  corresponds to a {\it fractional} $\frac{G^2 }{c^5}$ contribution
to scattering which  is determined (via the linear-response formula of Ref.~\cite{Bini:2012ji}) by
the leading-order {\it quadrupolar} fractional angular momentum radiated away during scattering,
equal to $ \frac{J^{\rm rad \,LO}_{\rm GR}}{J}= +\frac{16}{5}\frac{G^2 m_1 m_2 }{b^2 c^5} \,v$~\cite{Damour:1981bh} 
(see \cite{Damour:2020tta} for the exact $O(G^2)$ loss). 
The  $O(v^{-1})$ terms in $\Phi^{\rm sugra}_3$ and $\phi_3$ similarly correspond to
 leading-order {\it dipolar} fractional angular momentum scattering losses, which are 
$O\left( \frac{G^2 }{ c^3} \right)$ and  $O\left( \frac{(e_1e_2)^2 }{ c^3} \right)$, instead of $O\left( \frac{G^2 }{ c^5} \right)$. Using the couplings
 of extremal black holes  to the gauge, $A_\mu$,  $B_\mu$, and scalar, $\Phi, A, B$ 
 fields \cite{DiVecchia:2021ndb}, we indeed found a leading-order dipolar fractional angular momentum loss
during supergravity scattering  equal to: $ \left(\frac{J^{\rm rad}}{J}\right)^{\rm LO}_{\rm sugra}= +\frac{80}{3}\frac{G^2 m_1 m_2 }{b^2 c^3} \,v^{-1}$. The corresponding (Larmor-type) ED angular momentum
loss during scattering is equal to
$ \left(\frac{J^{\rm rad}}{J}\right)^{\rm LO}_{\rm ED}= -\frac{4}{3}\frac{e_1 e_2 }{b^2 c^3} \left( \frac{e_1}{m_1} - \frac{e_2}{m_2}\right)^2 \,v^{-1}$. Note the curious fact that the latter angular momentum loss
 is positive in the attractive ED case ($e_1 e_2 <0$), but {\it negative} in the repulsive ED case ($e_1 e_2 >0$). (see \cite{Saketh:2021sri} for  the exact $O((e_1e_2)^2)$ loss).

\bibliography{refs20231210.bib,local.bib}

\end{document}